\documentclass[pdflatex,sn-basic]{sn-jnl}

\usepackage{pgfplots}
\usepgfplotslibrary{statistics}
\usepackage{filecontents}

\usepackage{multirow}
\usepackage{booktabs}
\usepackage{rotating}
\usepackage[normalem]{ulem}
\usepackage{tabularray}
\usepackage{makecell}
\usepackage{colortbl}

\usepackage{graphicx}%
\usepackage{multirow}%
\usepackage{amsmath,amssymb,amsfonts}%
\usepackage{amsthm}%
\usepackage{mathrsfs}%
\usepackage[title]{appendix}%
\usepackage{xcolor}%
\usepackage{textcomp}%
\usepackage{manyfoot}%
\usepackage{booktabs}%
\usepackage{algorithm}%
\usepackage{algorithmicx}%
\usepackage{algpseudocode}%
\usepackage{listings}%

\usepackage{etoolbox}
\usepackage{ellipsis} 
\usepackage{xstring}
\usepackage{xparse} 
\AtBeginEnvironment{quote}{\itshape\small\vspace{-0.5em}}
\AfterEndEnvironment{quote}{\vspace{-0.5em}}

\newcommand{\textquote}[2][0]{ %
    \ifnum#1=1
        \textit{``[\dots] #2''}
    \else\ifnum#1=2
        \textit{``#2 [\dots]''}
    \else\ifnum#1=3
        \textit{``[\dots] #2 [\dots]''}
    \else
        \textit{``#2''}
    \fi\fi\fi
}

\usepackage[most]{tcolorbox}
\tcbset{colback=gray!10, colframe=gray!40!black, boxrule=0.4pt, arc=2pt, left=6pt, right=6pt, top=4pt, bottom=4pt, fonttitle=\bfseries}

\newtcolorbox{quoteBox}{colback=gray!5, colframe=white, sharp corners, boxrule=0pt, enhanced jigsaw, before skip=5pt, after skip=5pt, leftrule=2pt, rightrule=0pt, toprule=0pt, bottomrule=0pt, borderline west={2pt}{0pt}{gray!50!black}}

\newcommand{\keyword}[1]{\texttt{#1} }
\newcommand{\keywordN}[1]{\texttt{#1}}

\begin{document}

\title[What Learners Prefer to Motivate Their Learning]{Gamification with Purpose: What Learners Prefer to Motivate Their Learning}


\author*[1,2]{\fnm{Kai} \sur{Marquardt}}\email{kai.marquardt@kit.edu}
\equalcont{These authors contributed equally to this work.}

\author*[]{\fnm{Mona} \sur{Schulz}}\email{ugrwh@student.kit.edu}
\equalcont{These authors contributed equally to this work.}

\author*[1]{\fnm{Anne} \sur{Koziolek}}\email{koziolek@kit.edu}

\author*[1,2,3]{\fnm{Lucia} \sur{Happe}}\email{lucia.happe@kit.edu}
\equalcont{These authors contributed equally to this work.}
\affil*[1]{\orgdiv{Department of Informatics}, \orgname{Karlsruhe Institute of Technology}, \orgaddress{
\city{Karlsruhe}, 
\country{Germany}}
}

\affil*[2]{\orgdiv{Department for Interdisciplinary Didactics}, \orgname{Karlsruhe Institute of Technology}, \orgaddress{
\city{Kalsruhe}, 
\country{Germany}}
}

\affil*[3]{\orgdiv{Faculty of Business Management}, \orgname{University of Economics}, \orgaddress{
\city{Bratislava}, 
\country{Slovakia}}
}




\abstract{
This study investigates learners’ preferences for game design elements (GDEs) in educational contexts to inform the development of purpose-driven gamification strategies. It emphasizes a learner-centered approach that aligns gamification design with pedagogical goals, while mitigating risks such as the erosion of intrinsic motivation.
A systematic literature review was conducted to identify ten widely discussed GDEs. Visual prototypes representing each element were developed, and a best-worst scaling (BWS) survey with 125 participants was administered to elicit preference rankings. Qualitative feedback was also collected to uncover motivational drivers.
Learners consistently preferred GDEs that support learning processes directly—most notably progress bars, concept maps, immediate feedback, and achievements. Qualitative analysis revealed six recurring motivational themes, including visible progress, content relevance, and constructive feedback.
The findings suggest that learners value gamification elements that are meaningfully integrated with educational content and support intrinsic motivation. Purpose-aligned gamification should prioritize tools that visualize learning progress and provide actionable feedback, rather than relying solely on extrinsic incentives.
}





\keywords{Educational gamification, game design elements, learning motivation, learner preferences, best-worst scaling, digital learning environments}



\maketitle
\section{Introduction}
Gamification, the integration of game design elements (GDEs) into non-game contexts, has emerged as a widely adopted strategy in educational settings to enhance student motivation, engagement, and learning outcomes \citep{deterding2011game, kalogiannakis2021gamification, dehghanzadeh2023gamification}. By infusing educational experiences with elements such as points, feedback, or levels, gamification aims to transform often passive learning environments into interactive and motivating ones. This transformation is particularly valuable in digital and hybrid learning contexts, where sustaining learner attention can be challenging.

The interest in gamification is especially pronounced in fields such as STEM and computer science education, which continue to grapple with issues of learner disengagement and underrepresentation of diverse student populations \citep{albusays2021diversity, happe2022frustrations, wang2017diversity}. In such areas, gamification holds promise as a tool to bridge engagement gaps and promote inclusive learning. However, the design of gamified learning environments is not straightforward. Decisions about which GDEs to implement—and how to implement them—are often made without a clear understanding of learners' actual preferences or the motivational mechanisms these elements support.

Much of the existing research has focused on a narrow subset of gamification elements, particularly the widely known PBL triad: points, badges, and leaderboards \citep{dehghanzadeh2023gamification, klock2020tailored}. While these elements have proven useful in various domains, the field lacks comprehensive empirical studies exploring a broader set of GDEs, especially in educational contexts. Moreover, studies that include user preferences typically evaluate elements in isolation rather than in combination, overlooking potential synergies between GDEs. The optimal number, combination, and context-sensitive application of these elements remain open questions \citep{khaldi2023gamification, manzano2021between, nadi2022gamification}.

Furthermore, the indiscriminate use of gamification can backfire. Overemphasis on extrinsic rewards, such as points or virtual currency, may undermine intrinsic motivation and lead to superficial engagement \citep{almeida2023negative}. A more sustainable and pedagogically grounded approach requires understanding which GDEs learners themselves perceive as motivating and why. This calls for a shift toward purpose-driven gamification—design strategies that align GDEs with educational goals and learners' motivational needs.

This study responds to that call by investigating learner preferences for ten game design elements in educational contexts. We conducted a systematic literature review to identify both widely used (e.g., points, badges, leaderboards) and less-explored elements (e.g., storytelling, concept maps). While concept maps are typically used as instructional scaffolds rather than game elements, we explore their potential as motivational tools for visualizing progress within gamified learning contexts—a novel interpretation that reflects learners’ desire for content-aligned feedback. The latter, particularly concept maps, are rarely considered gamification elements but may offer pedagogically valuable means of visualizing progress and structuring learning. To evaluate preferences, we developed visual prototypes of each GDE and conducted a survey using the best-worst scaling (BWS) method, a technique that forces participants to make meaningful trade-offs and rank elements more reliably. We also analyzed open-ended responses to understand the motivational drivers behind learners’ choices. We hypothesize that game design elements which visualize progress—such as progress bars, achievements, and concept maps—may be perceived as more motivating, particularly because they connect learners’ actions with visible growth. This assumption guides our inclusion of both conventional and less-explored GDEs that serve this function.

In addition to identifying preferred GDEs, we examined whether preferences vary according to learner traits such as age, gender, and enjoyment of video games. This analysis helps determine whether gamification strategies should be tailored to specific user groups. We also explored the reasons learners found certain GDEs more effective for motivating their learning, aiming to extract underlying themes or motivational categories that can guide future design. Notably, the results not only highlight individual GDE preferences but also indicate that learners evaluate gamified learning experiences in terms of meaningful progress, feedback, and alignment with their personal goals—supporting the need for more targeted and learner-centered gamification.

Our study is guided by the following research questions:

\begin{itemize}
\item \textbf{RQ1:} What are the most preferred GDEs by learners for enhancing their learning motivation, and do these preferences vary based on user traits (e.g., age, video game enjoyment, and gender)?
\item \textbf{RQ2:} What combinations of GDEs do learners most prefer?
\item \textbf{RQ3:} Why do learners perceive specific GDEs as effective for motivating learning?
\end{itemize}

By answering these questions, this study contributes empirical evidence for learner-centered, purpose-driven gamification in education. The results offer practical guidance for educators, instructional designers, and developers seeking to create more engaging and effective learning environments. In particular, the findings highlight the potential of lesser-studied elements like concept maps and the importance of designing GDE combinations that reflect learner preferences. This moves the field beyond the traditional PBL triad and opens up new avenues for personalized and pedagogically meaningful gamification. These insights stimulate discussions on innovative approaches to incorporating gamification that enhance learner satisfaction, cater to individual needs, and ultimately contribute to the creation of more effective technology-supported learning environments within the educational system.

\section{Background}
Gamification stems from the concept of leveraging the motivational aspects of games in various real-world scenarios. Given its multifaceted nature, a comprehensive and universally accepted definition of gamification remains elusive. The most widely accepted definition, proposed by Deterding et al., characterizes gamification as the integration of game design elements (GDEs) into non-game contexts \citep{deterding2011game}. The term game design elements (GDEs) itself is maybe even less rigidly defined and is generally described as \textquote[0]{elements that are characteristic to games} \citep[p.12]{deterding2011game}. Over time, various taxonomies have been proposed in an effort to provide clearer guidelines and reduce the ambiguity associated with gamification's definition \citep{toda2019analysing, azouz2018gamification}.

The aim of gamification is to increase motivation and cause behavioral changes through a more playful environment \citep{sailer2017gamification}. Gamification finds application across a broad spectrum, including areas like fitness and finance. Notably, in recent years, it has gained increased attention within the are of education due to its potential to make educational environments more engaging \citep{dicheva2020badges,Venter2019, Dicheva2021MoEn}, improve motivation \citep{Venter2019, Dicheva2021MoEn}, and enhance learning outcomes \citep{dicheva2020badges}. Game-based learning and serious games are closely related fields of study, but it's essential to differentiate them from gamification. Contrary to gamification, these approaches entail the integration of non-game elements into a gaming context to deliver educational content through interactive gameplay \citep{krath2021revealing}.
Gamification involves the integration of playful elements into tasks or activities. These tasks are designed to be entertaining \citep{sailer2017gamification} and typically feature a combination of GDEs, for example those from the common triad of points, badges, and leaderboards (PBL). Research suggests that using isolated GDEs may be less effective or even yield negative outcomes \citep{manzano2021between}. However, the specific combinations of GDEs that are most effective in educational gamification contexts remain unclear \citep{nadi2022gamification, khaldi2023gamification}.

In addition to the documented benefits of gamification, its potential negative impacts have been a subject of debate within the literature \citep{almeida2023negative}. These adverse effects arise when the utilization of these systems is predominantly driven by extrinsic motivation, potentially compromising the overall learning experience and impeding learning outcomes \citep{almeida2023negative, hakulinen2015effect, reid2015digital, Venter2019, sailer2017gamification}. In educational settings, the effectiveness of gamification hinges on its ability to tap into intrinsic motivation \citep{buckley2016gamification}.
Recently, a new dimension in gamification research, known as tailored gamification, has garnered attention. This approach acknowledges that individuals learn differently and are motivated by distinct factors, emphasizing the importance of customized gamification strategies \citep{klock2020tailored, Rodrigues2021personalization, el2020user}.

In the design of educational systems, the primary goal of gamification should be to enrich the learning process by amplifying learning motivation. This is primarily achieved by nurturing learners' intrinsic motivation to actively engage with the content \citep{deci2001extrinsic}. So we understand learning motivation as any intrinsic drive, that can be technology-supported by gamification, that encourages learners to engage with the educational material. While the field of gamification in education has reached growing attention in research, numerous questions remain unanswered concerning the optimal utilization of gamification to intrinsically enhance learning motivation \citep{khaldi2023gamification, nadi2022gamification}. This study aims to provide insights into learners' attitudes toward newly defined gamification elements closely aligned with the learning content, shedding light on this ongoing inquiry.

\section{Research Method}
To understand what game design elements (GDEs) and combination of GDEs show the highest potential to increase learning motivation of learners, we conducted a three-stage study. In the first stage, we conducted a systematic literature review to identify relevant GDEs. Based on results from the literature review, we selected ten GDEs and created visual prototypes for each GDE. In the third stage, we conducted a survey study with a best-worst scaling (BWS) approach to assess the relative importance of the ten GDEs.

\subsection{Selection of Game Design Elements}
In this section, we provide a short overview of our selection process.
Depending on the source there are more than 100 different GDEs, and these elements may not consistently adhere to the same categorization standards \citep{voit2020towards, toda2019analysing}. To get an overview of potentially effective GDEs we performed a systematic literature review. We searched various literature databases like ACM Digital Library, ERIC, IEEE Xplore, JSTOR and SpringerLink with combinations of the keywords \texttt{gamif*}, \texttt{intrinsic}, \texttt{motivation}, \texttt{engagement}, \texttt{element}, \texttt{design}, and \texttt{educat*}. Following the initial search, we identified a total of 348 potentially relevant papers. Subsequent to removing duplicates and conducting a thorough examination based on our predefined inclusion and exclusion criteria, we narrowed down our selection to 18 papers. Through a snowballing procedure, by still following interesting references, three additional papers were included, leading to a final data set of 21 papers.
Within this subset, we identified 28 different GDEs that exhibited potential effectiveness in enhancing learning motivation.
Based on the findings from our literature review, we identified nine GDEs that matched our interest in studying their potential to enhance learning motivation: achievements, badges, progress bars, feedback, storytelling, leaderboards, points, virtual currency, and avatars.
s the tenth GDE we included the concept map in our study. We made this selection based on our interest and vision for this element. Concept maps can be categorized as a type of progression GDE \citep{toda2019analysing} and are related to other recognized terms such as knowledge maps or progression graphs. However, there is a very limited number of studies available that examine this type of GDE \citep{klock2020tailored, borges2016link}.
\begin{table}[]
    \centering
\begin{tabular}{llp{8cm}}
\toprule
    Element & Count & Source \\
\midrule
    Badges (BA) & 13 & \citep{Dicheva2021MoEn,Rodrigues2021personalization,Jesus,Thamrongrat,dicheva2020badges,roubi2019towards,Venter2019,Karra2019,Jesus2019,Schwarzmann,hakulinen2015effect,reid2015digital,hallifax2019factors} \\
    Points (PO) & 9 & \citep{Dicheva2021MoEn,Rodrigues2021personalization,Jesus,Thamrongrat,ccubukccu2017gamification,roubi2019towards,Jesus2019,sailer2017gamification,hallifax2019factors} \\
    Avatar (AV) & 7 & \citep{Dicheva2021MoEn,Jesus,Qiu2021,Venter2019,Karra2019,Jesus2019,Schwarzmann} \\
    Leaderboards (LB) & 6 & \citep{Dicheva2021MoEn,Rodrigues2021personalization,Jesus,ccubukccu2017gamification,Schwarzmann,hallifax2019factors} \\
    Achievements (AC) & 5 & \citep{hallifax2019factors,Karra2019,ccubukccu2017gamification,Jesus,botte2020motivation} \\
    Progress Bar (PB) & 4 & \citep{Karra2019,Rodrigues2021personalization,Schwarzmann,Venter2019} \\
    Virtual Currency (VC) & 4 & \citep{Dicheva2021MoEn,Jesus,Dicheva2022,Jesus2019} \\
    Feedback (FB) & 2 & \citep{Rodrigues2021personalization,Karra2019} \\
    Storytelling (ST) & 1 & \citep{Schwarzmann} \\
\bottomrule
\end{tabular}
 \caption{Frequencies and sources of GDEs from the literature review}
    \label{tab:gdes_slr}
\end{table}


For the survey, we created visual prototypes for all ten selected GDEs to provide participants with illustrative examples for reference.
During the design process of these elements, our aim was to achieve a balance between providing visualizations that are as general as possible, yet including all the necessary and characteristic details unique to each element and aligned with our vision of their real-world implementation. All visual prototypes as well as their descriptions can be found in the appendix \ref{appendix:prototypes}.

\subsection{Best-Worst Scaling (BWS)}
To identify GDEs that motivate learners in a digital learning context the most, we use a best-worst scaling (BWS) approach \citep{louviere2013introduction}, which requires our participants to choose the most preferred and least preferred options from a set of GDEs, which we will refer as a choice set. 
Following the guidelines by Orme \citep{orme2005accuracy} we used a total of 15 choice sets each consisting of four different GDEs (see appendix \ref{appendix:choicesets}). Each element is shown exactly six times in total. Participants then iteratively choose the most and least preferred GDE out of the set of four presented GDEs.

Best-Worst Scaling (BWS) has been effectively employed in various educational contexts \citep{huybers2014student, schoebel2016agony, melo2022assess} and offers several advantages. It presents an easy-to-follow task for participants, provides clear ranking information, operates within a scale-free framework, thereby reducing the impact of response styles on mean values and variances \citep{schoebel2016agony, flynn2007best}, and it helps minimize various response biases \citep{lee2007measuring}. BWS was selected over Likert scales to minimize central tendency bias and force more nuanced trade-offs between elements. 

\subsection{Survey Study}
The goal of the survey study was to identify the most preferred GDEs and combinations of GDEs participants would choose as being most effective in increasing their learning motivation.
At the beginning of the survey, a short introduction to the research project was given, where the participants were instructed to envision these GDEs within a setting, where they use digital learning environments (such as online courses) to learn new content. Then all ten GDEs with their prototypes and descriptions were presented to the participants. After this section, the main part of the survey began with the BWS approach comprising 15 rounds of choice sets. Each choice set consisted of four GDEs and participants were required to select one GDE as their most preferred and one as their least preferred choice from the set of four elements. Throughout these rounds, participants had continuous access to the prototypes and their descriptions.
After the BWS part, participants were asked to freely choose their preferred combination of GDEs (out of the 10) and to explain which of the chosen GDEs would motivate them the most and why. 
Subsequent questions covered demographic information such as gender (female, male, non-binary, self-described) and age as well as inquiries regarding their enjoyment of playing video games and preferred game genres.
The survey was conducted with Google Forms and shared via personal contacts (snowballing) and social media.

\subsection{Participants}
\begin{figure}
    \centering
        \begin{tikzpicture}
        \begin{axis}[
            ybar,
            width=0.5\textwidth,
            height=5cm,
            xlabel={\footnotesize Age},
            ymin=0,
            x tick label style={rotate=0, anchor=center},
            legend pos=north east,
            bar width=2pt, 
            ymajorgrids
        ]
    
        \addplot [fill=blue!30] coordinates {
            (15, 1)
            (16, 6)
            (17, 7)
            (18, 7)
            (19, 5)
            (20, 3)
            (21, 14)
            (22, 8)
            (23, 14)
            (24, 13)
            (25, 4)
            (26, 5)
            (28, 3)
            (29, 1)
            (31, 1)
            (33, 2)
            (40, 1)
            (43, 1)
            (46, 1)
            (47, 1)
            (51, 1)
            (52, 3)
            (53, 3)
            (54, 2)
            (55, 1)
            (56, 1)
            (57, 2)
            (59, 1)
            (64, 1)
        };
        
        \end{axis}
        \end{tikzpicture}
    \caption{Age distribution of participants}
    \label{fig:age-distribution}
\end{figure}
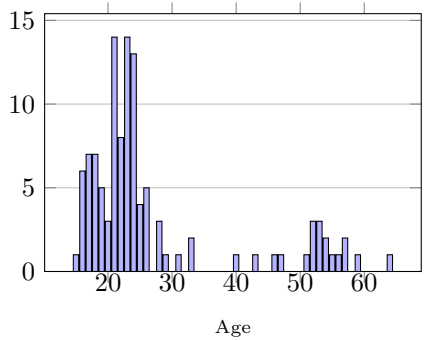

In total, we had 125 participants responding to our survey, comprising 65 (53\%) males, 56 (44\%) females, one (1\%) non-binary, and three (2\%) participants who did not indicate their gender. Participant ages ranged from 15 to 64 years, with the majority falling between 20 and 30 years old (mean 27.02, median 23, see Figure \ref{fig:age-distribution}).

\subsection{Analysis}
For the BWS analysis, we employed a counting approach to determine the ranking positions of the GDEs. Following the established procedure in related literature \citep{schoebel2016agony, orme2005accuracy, berger2021gamification, schmidt2019users}, we tallied the frequency with which each GDE was chosen as the best and worst, and based on these results, we calculated standardized means and standard deviations for each element.
To further validate the results obtained from the counting analysis, we employed a logic regression model, which provides a more detailed explanation of the relative ranking of the GDEs \citep{flynn2007best, lipovetsky2014best}
For the combination analysis, we calculated frequencies to identify the most preferred combinations of GDEs and also examined correlations to gain insights into the relationships between pairs of elements.

In our qualitative analysis of participants' responses to the open-ended survey question about why they selected specific GDEs as the most effective for enhancing their learning motivation, we employed an iterative coding approach to establish categories of motivation drivers \citep{saldana2013coding}. Initially, we formalized keywords for each response, and then we iteratively refined and merged these keywords to create categories related to learning motivation. To ensure consistency and comparability, we translated all German responses into English, correcting any grammar and spelling errors in the process.

For our statistical analyses, we used the software tools Excel, SPSS, and the package \emph{bwsTools} in R for the regression analysis.

\section{Findings}
In this section, we present our findings organized according to the three research questions. We will begin with the results of the BWS analysis, followed by the results of the combination analysis, and conclude with the findings from the qualitative analysis, highlighting the various categories of motivational drivers associated with the different GDEs.

\subsection{Results of the Best-Worst-Scaling}
\begin{table*}[]
  \centering
  \resizebox{\linewidth}{!}{%
\begin{tblr}{
  row{1} = {t},
  row{2} = {t},
  row{3} = {t},
  cell{1}{2} = {c=8}{},
  cell{1}{11} = {c=4}{},
  cell{1}{15} = {r=3}{},
  cell{2}{2} = {c=3}{},
  cell{2}{5} = {c=3}{},
  cell{2}{13} = {c=2}{},
  hline{1,4,14} = {-}{},
  hline{2} = {2-9,11-14}{},
  hline{3} = {2-7,13-14}{},
}
Element          & Counting Analysis &      &       &                  &                   &        &          &       &  & Regression Analysis    &       &         &        & {Choice\\prob-\\ability} & Rank \\
                 & Counts            &      &       & Proportions      &                   &        & {Std.\\Mean} & SD    &  & {Coeff.}            & SD    & 95\%-CI &        &                          &      \\
                 & Total             & Best & Worst & {Best\\in total} & {Worst\\in total} & Diff   &          &       &  &                        &       & Lower   & Upper  &                          &      \\
Progress
  Bar   & 600               & 363  & 49    & 0.605            & 0.082             & 0.523  & 0.419    & 0.403 &  & 1.162                  & 0.068 & 1.029   & 1.295  & 0.254                    & 1    \\
Achievements         & 600               & 241  & 59    & 0.402            & 0.098             & 0.303  & 0.243    & 0.369 &  & 0.626                  & 0.061 & 0.508   & 0.745  & 0.148                    & 2    \\
Feedback      & 600               & 263  & 103   & 0.438            & 0.172             & 0.267  & 0.213    & 0.500 &  & 0.547                  & 0.060 & 0.429   & 0.664  & 0.137                    & 3    \\
Concept Map     & 600               & 248  & 131   & 0.413            & 0.218             & 0.195  & 0.156    & 0.555 &  &  0.395 & 0.059 & 0.280   & 0.510  & 0.118                    & 4    \\
Points           & 600               & 185  & 107   & 0.308            & 0.178             & 0.130  & 0.104    & 0.380 &  & 0.261                  & 0.058 & 0.147   & 0.376  & 0.103                    & 5    \\
Badges           & 600               & 154  & 147   & 0.257            & 0.245             & 0.012  & 0.009    & 0.449 &  & 0.023                  & 0.058 & -0.090  & 0.137  & 0.081                    & 6    \\
Leaderboards     & 600               & 137  & 240   & 0.228            & 0.400             & -0.172 & -0.137   & 0.525 &  & -0.347                 & 0.059 & -0.462  & -0.232 & 0.056                    & 7    \\
Storytelling      & 600               & 138  & 248   & 0.230            & 0.413             & -0.183 & -0.147   & 0.546 &  & -0.371                 & 0.059 & -0.486  & -0.256 & 0.055                    & 8    \\
Virtual Currency & 600               & 96   & 371   & 0.160            & 0.618             & -0.458 & -0.367   & 0.560 &  & -0.990                  & 0.065 & -1.118  & -0.863 & 0.029                    & 9    \\
Avatar           & 600               & 50   & 420   & 0.083            & 0.700             & -0.617 & -0.493   & 0.451 &  & -1.439                 & 0.073 & -1.583  & -1.295 & 0.019                    & 10   
\end{tblr}
}
  \caption{Results of the counting and regression analysis for the BWS ranking}
  \label{t_bws}
\end{table*}

Table \ref{t_bws} shows the results of the counting analysis. The standardized mean (Std.Mean) was calculated by the difference between the number of times the respective GDE was preferred the most and the least, and then divided by the product of the number each GDE appeared in a choice set (6) and the number of participants (125). For the element progress bar, the calculation would be for example: (363-49)/(6*125)= 0,4187. The resulting scale ranges from -1 to 1, with values closer to 1 indicating higher preferences and values closer to -1 indicating lower preferences. 

The results according to standardized means show that the progress bar emerged as the most preferred GDE, exhibiting a significant lead over the next most preferred elements, which include achievements, feedback as the top 3 ranked GDES, and concept map, and points as the top 5 ranked GDEs. These are followed by badges, leaderboards, and storytelling. The least preferred GDEs was the avatar, followed by virtual currency.
Examining the raw data, as indicated by the columns for frequencies of best and worst selections of a GDE, further highlights these findings. The progress bar was chosen as the most preferred alternative much more frequently than any other GDE. In total, it was chosen as the best alternative in more than 60\% of its occurrences (n=363), which corresponds to a 70\% higher frequency than the next most often chosen GDE, which is feedback (n=263). Following feedback, concept map (n=248) and achievements (n=241) were the next most frequently chosen GDEs. The other GDEs had a more substantial gap between them. Avatar and virtual currency were also consistently the most often chosen as the least preferable alternatives, once again with a significant margin.
The regression analysis verifies the results from the counting analysis as it shows the same ranking.

\begin{table*}[]
  \centering
  \resizebox{\linewidth}{!}{%
    \begin{tabular}{ccccccccc}
    \toprule
    \multirow{2}{*}{Rank} & \multirow{2}{*}{Overall} & \multicolumn{2}{c}{Gender} & \multicolumn{2}{c}{Video Game Enjoyment} & \multicolumn{3}{c}{Age}                                 \\
                          &                          & Female                      & Male                                      & Less Enjoyment            & Enjoyment & $<20$ & $20-26$ & $>26$ \\
    \midrule
    1                     & PB                       & PB                          & PB                                        & PB                       & PB        & PB  & PB    & PB  \\
    2                     & AC                       & FB                          & AC                                        & FB                       & AC        & PO  & AC    & CM  \\
    3                     & FB                       & AC                          & CM                                        & AC                       & FB        & AC  & FB    & AC  \\
    4                     & CM                       & CM                          & FB                                        & PO                       & CM        & BA  & CM    & FB  \\
    5                     & PO                       & PO                          & PO                                        & CM                       & PO        & CM  & PO    & PO  \\
    6                     & BA                       & ST                          & BA                                        & BA                       & BA        & FB  & BA    & BA  \\
    7                     & LB                       & BA                          & LB                                        & ST                       & LB        & LB  & ST    & ST  \\
    8                     & ST                       & LB                          & ST                                        & LB                       & ST        & VC  & LB    & LB  \\
    9                     & VC                       & VC                          & VC                                        & AV                       & VC        & ST  & VC    & VC  \\
    10                    & AV                       & AV                          & AV                                        & VC                       & AV        & AV  & AV    & AV  \\
    \midrule
    n                    & 125                       & 56                          & 65                                        & 44                       & 80        & 26  & 61    & 26  \\
    \bottomrule
    \end{tabular}
    }
  \caption{Comparison of BWS ranking results by different user traits}
  \label{t_bws_ut}
\end{table*}
To investigate whether there are preference differences based on user traits, we performed also a counting analysis differentiated by gender, video game enjoyment, and age. Regarding the analysis of video game enjoyment, we categorized our sample based on their responses to the survey question, \textquote{Do you enjoy playing video games?}. Responses \textquote{Not at all}, \textquote{Not really}, \textquote{Neutral} were grouped as \texttt{Less Enjoyment} and responses \textquote{Somewhat} and \textquote{Very much} were grouped as \texttt{Enjoyment}. We acknowledge that grouping \textquote{Neutral} responses with \texttt{Less Enjoyment} may oversimplify participants’ actual preferences. Future analyses could benefit from a three-group categorization (low, neutral, high) to capture more nuanced trends. In this version, we chose the two-group split to enable clearer statistical comparisons. Due to the limited diversity in our sample regarding age, we decided to generalize the age categories by learning stages. Specifically, we categorized participants as follows: age 15-20 as secondary school to early tertiary education, age 20-26 as tertiary education (e.g. university, vocational training), and age >26 as employed individuals.

Table \ref{t_bws_ut} presents the results of the GDEs ranking based on the counting analysis differentiated by user traits. The findings indicate only minor differences among different user traits. For most user traits, the top four ranked GDEs consisted of different combinations of the overall ranking, with the progress bar consistently being the highest-ranked element.
There were no significant gender differences, with male participants showing a slightly higher preference for achievements and leaderboards, while female participants showed a slightly higher preference for feedback and storytelling, compared to each other.
Regarding video game enjoyment, no major differences were observed, except for minor rearrangements. Standardized means for participants indicating high video game enjoyment suggested that they have clearer preferences (ranging from 0.18 to 0.45 among the top 4 elements) compared to those less interested in video games (ranging from 0.11 to 0.37 among the top 4 elements).
The age-related analysis indicated that younger learners (age 15-19) had a higher preference for the GDE points (mean=0.16), compared to the age groups 20 to 26 (mean=0.06) and older than 26 (mean=0.05). On the other hand, the GDEs concept map and feedback seemed to be less important for the younger group (mean$_{CM,<20}$=0.00, mean$_{FB,<20}$=-0.03) compared to older participants (mean$_{CM,20-26}$=0.12, mean$_{CM,>26}$=0.28, mean$_{FB,20-26}$=0.15, mean$_{FB,>26}$=0.22). 
Overall, while there were some differences based on user traits, the general pattern of GDE preferences remained relatively consistent across different groups.

\subsection{Results of the Combination Analysis}
\begin{table}[]
  \centering
  \begin{tabular}{llc}
    \toprule
    Element          & Frequency & Rank BWS      \\
    \midrule
    Progress Bar     & 89 (70.63\%)  & 1  \\
    
    Achievements     & 81 (64.29\%)  & 2  \\
    
    Concept Map      & 74 (58.73\%)  & 4  \\
    
    Feedback         & 73 (57.94\%)  & 3  \\
    
    Points           & 64 (50.79\%)  & 5  \\
    
    Badges           & 43 (34.13\%)  & 6  \\
    
    Leaderboard      & 43 (34.13\%)  & 7  \\
    
    Storytelling     & 42 (33.33\%)  & 8  \\
    
    Avatar           & 27 (21.43\%)  & 10 \\
    
    Virtual Currency & 20 (15.87\%)  & 9  \\
    
    None of Them     & 2  (1.59\%)   & -  \\
    \bottomrule
  \end{tabular}
  \caption{Frequency of GDEs in bundles}
  \label{t_bundle_freq_bestranked}
\end{table}

\begin{table*}[]
    \centering
    \resizebox{\linewidth}{!}{%
\begin{tabular}{lccccccccccl}
\toprule
    \textbf{Number of Elements in Bundle} & \textbf{1} & \textbf{2} & \textbf{3} & \textbf{4} & \textbf{5} & \textbf{6} & \textbf{7} & \textbf{8} & \textbf{9} & \textbf{10} &  \textbf{$\sum$} \\
\midrule
    Number of Responses (total) & 6 & 7 & 28 & 20 & 28 & 18 & 8 & 6 & 0 & 2 & 123 (100.0\%) \\
\midrule
    Progress Bar & 0 & 2 & \textbf{16} & \textbf{13} & \textbf{26} & 16 & 8 & 6 & 0 & 2 & 89 (72.4\%) \\
    Achievements & 2 & 2 & 12 & 9 & 25 & \textbf{18} & 5 & 6 & 0 & 2 & 81 (65.9\%) \\
    Concept Map & 2 & 2 & 12 & \textbf{13} & 20 & 11 & 7 & 5 & 0 & 2 & 74 (60.2\%) \\
    Feedback & 0 & 1 & 9 & 12 & 20 & 15 & 8 & 6 & 0 & 2 & 73 (59.3\%) \\
\midrule
    At least one of the best four elements& 4 & 5 & 25 & 19 & 28 & 18 & 8 & 6 & 0 & 2 & 115 (93.5\%) \\
    At least two of the best four elements& - & 2 & 18 & 17 & 28 & 18 & 8 & 6 & 0 & 2 & 99 (80.5\%) \\
    At least three of the best four elements & - & - & 6 & 9 & 23 & 14 & 8 & 6 & 0 & 2 & 68 (59.1\%) \\
    All four elements & - & - & - & 2 & 12 & 10 & 4 & 5 & 0 &2 & 35 (28.5\%) \\ 
\bottomrule
\end{tabular}
}
\caption{Combination and frequency of best-ranked GDEs in bundles}
    \label{tab:topbundle}
\end{table*}

On average, participants selected 4.45 GDEs in bundles at the survey question \textquote{If you were free to choose between the previous game design elements, which would you use together in online courses to increase your learning motivation?}. Table \ref{t_bundle_freq_bestranked} displays the frequency of each GDE being chosen as part of a bundle, and Table \ref{tab:topbundle} provides a more detailed breakdown for different bundle sizes.

The order of the most frequently chosen elements closely aligns with the ranking from the counting analysis, with the only exceptions being that concept map switched positions with feedback, and avatar switched positions with virtual currency. The progress bar element remains the most commonly chosen element, included in the bundles of over 70\% of participants. Achievements are the second most frequently chosen element in GDE bundles, followed by concept map and feedback.
For bundles of size three and five, the progress bar was the most frequently chosen element (see Table \ref{tab:topbundle}). For bundles of size three, the top choices were the progress bar and concept map. For bundles of size six, achievements were most often chosen.

Overall, more than 90\% of participants included at least one of these top four preferred GDEs in their bundles, with over 80\% including more than two of them in their selections. Even about 60\% of participants included at least three of the top four elements in their bundles. Approximately 30\% included all four of the top elements in their bundle selection, rising to 42\% when considering only participants who chose bundle sizes of 4 and higher.

We also conducted a correlation analysis to identify elements that were more likely to be chosen together. There were only moderate correlations found between achievements and progress bar ($r = .382, p < .01$), achievements and badges ($r = .322, p < .01$), feedback and progress bar ($r = .323, p < .01$), and virtual currency and avatar ($r = .345, p < .01$).

A differentiated analysis for different user traits did not reveal any major differences, except for the younger age groups, where points were chosen significantly more frequently compared to the other age groups. For participants aged 19 and younger, points were the most often chosen element in bundles, while for the other age groups, the GDE points ranked fifth.

\subsection{Results from the Qualitative Analysis of Motivational Drivers}

\begin{table*}[]
    \resizebox{\linewidth}{!}{%
    \begin{tblr}{
  row{1} = {b},
  row{8} = {c},
  row{9} = {c},
  cell{1}{2} = {c},
  cell{1}{3} = {c},
  cell{1}{4} = {c},
  cell{1}{5} = {c},
  cell{1}{6} = {c},
  cell{1}{7} = {c},
  cell{1}{8} = {c},
  cell{1}{9} = {c},
  cell{1}{10} = {c},
  cell{1}{11} = {c},
  cell{2}{2} = {c},
  cell{2}{3} = {c},
  cell{2}{4} = {c},
  cell{2}{5} = {c},
  cell{2}{6} = {c},
  cell{2}{7} = {c},
  cell{2}{8} = {c},
  cell{2}{9} = {c},
  cell{2}{10} = {c},
  cell{2}{11} = {c},
  cell{2}{12} = {c},
  cell{3}{2} = {c},
  cell{3}{3} = {c},
  cell{3}{4} = {c},
  cell{3}{5} = {c},
  cell{3}{6} = {c},
  cell{3}{7} = {c},
  cell{3}{8} = {c},
  cell{3}{9} = {c},
  cell{3}{10} = {c},
  cell{3}{11} = {c},
  cell{3}{12} = {c},
  cell{4}{2} = {c},
  cell{4}{3} = {c},
  cell{4}{4} = {c},
  cell{4}{5} = {c},
  cell{4}{6} = {c},
  cell{4}{7} = {c},
  cell{4}{8} = {c},
  cell{4}{9} = {c},
  cell{4}{10} = {c},
  cell{4}{11} = {c},
  cell{4}{12} = {c},
  cell{5}{2} = {c},
  cell{5}{3} = {c},
  cell{5}{4} = {c},
  cell{5}{5} = {c},
  cell{5}{6} = {c},
  cell{5}{7} = {c},
  cell{5}{8} = {c},
  cell{5}{9} = {c},
  cell{5}{10} = {c},
  cell{5}{11} = {c},
  cell{5}{12} = {c},
  cell{6}{2} = {c},
  cell{6}{3} = {c},
  cell{6}{4} = {c},
  cell{6}{5} = {c},
  cell{6}{6} = {c},
  cell{6}{7} = {c},
  cell{6}{8} = {c},
  cell{6}{9} = {c},
  cell{6}{10} = {c},
  cell{6}{11} = {c},
  cell{6}{12} = {c},
  cell{7}{2} = {c},
  cell{7}{3} = {c},
  cell{7}{4} = {c},
  cell{7}{5} = {c},
  cell{7}{6} = {c},
  cell{7}{7} = {c},
  cell{7}{8} = {c},
  cell{7}{9} = {c},
  cell{7}{10} = {c},
  cell{7}{11} = {c},
  cell{7}{12} = {c},
  cell{8}{1} = {r},
  cell{9}{1} = {r},
  cell{10}{1} = {c=12}{},
  hline{1,10} = {-}{.75pt},
  hline{2-8} = {-}{},
}
{\textbf{CATEGORY}\\\textbf{\textit{Learning motivation through…}}} & \begin{sideways}\textbf{Progress Bar}\end{sideways} & \begin{sideways}\textbf{Concept Map}\end{sideways} & \begin{sideways}\textbf{Achievements}\end{sideways} & \begin{sideways}\textbf{Feedback}\end{sideways} & \begin{sideways}\textbf{Badges}\end{sideways} & \begin{sideways}\textbf{Storytelling}\end{sideways} & \begin{sideways}\textbf{Leaderboard}\end{sideways} & \begin{sideways}\textbf{Points}\end{sideways} & \begin{sideways}\textbf{Virtual Currency}\end{sideways} & \begin{sideways}\textbf{Avatar}\end{sideways} & $\sum_{cat}$\\
{VISUAL PROGRESSION\\\textit{… making learning progress and success distinctly~visible}} & \textbf{15} & 5 & \textbf{6} & - & \textbf{8} & - & - & 2 & - & - & \textbf{23}\\
{EFFECTIVE LEARNING\\\textit{… providing strategies that make~learning more effective}} & 1 & 3 & 1 & \textbf{12} & 2 & \textbf{4} & - & 1 & 1 & - & \textbf{20}\\
{GOALS\\\textit{… clear reachable goals that keep the learner engaged}} & 6 & 2 & \textbf{8} & - & \textbf{7} & - & - & - & 1 & - & \textbf{19}\\
{STRUCTURE\\\textit{… providing a clear structure and making connections visible}} & 4 & \textbf{13} & 1 & - & 1 & 1 & - & - & - & - & \textbf{18}\\
{SHOWCASE\\\textit{… social comparison and competition}} & - & - & 1 & - & 1 & - & \textbf{10} & 1 & 2 & - & \textbf{13}\\
{RELEVANCE\\\textit{… meaningful and relatable learning contexts}} & - & 1 & 2 & - & - & \textbf{5} & - & - & 1 & 1 & \textbf{8}\\
~$\sum_{gde}$ & \textbf{24} & \textbf{20} & \textbf{16} & \textbf{15} & \textbf{14} & \textbf{13} & \textbf{12} & \textbf{10} & \textbf{6} & \textbf{2} & \textbf{88}\\
~$\sum_{gde}'$ & \textit{16} & \textit{17} & \textit{10} & \textit{11} & \textit{9} & \textit{7} & \textit{7} & \textit{3} & \textit{4} & \textit{1} & \textit{66}
\\
{\footnotesize $\sum_{gde}$ responses related to the respective GDE\\$\sum_{gde}'$ responses related to the respective GDE including only those that could be assigned to a category\\$\sum_{cat}$ total responses associated with the respective category} &  &  &  &  &  &  &  &  &  &  & 
\end{tblr}
}
    \caption{Results from the qualitative analysis of the survey question \textquote{Which game design elements in the given scenario would motivate you the most and why?}}
    \label{tab:qualitative-analysis}
\end{table*}

In this section, we present the results of the qualitative analysis based on responses to the survey question, \textquote{Which game design elements in the given scenario would motivate you the most and why?}. We received a total of 88 responses to this question.
Table \ref{tab:qualitative-analysis} reveals that the GDEs progress bar (n=24) and concept map (n=20) were each mentioned by approximately a quarter of all responses as the most motivating GDEs. Following closely, the GDEs achievements (n=16) and feedback (n=15) were also frequently cited. In contrast, virtual currency (n=6) and avatar (n=2) were the least favored GDEs in terms of their perceived value for enhancing learning motivation. These findings align with the results obtained from the combination analysis.
While the frequency analysis of mentioned GDEs offers valuable insights into their relative importance, the participants' provided reasons for selecting specific GDEs allow for a deeper understanding of their motivational drivers.

Through our systematic coding approach, we identified six categories of learning motivation that summarize the reasons participants provided for selecting specific GDEs as the most effective for motivating them: \keywordN{VISUAL PROGRESSION}, \keywordN{EFFECTIVE LEARNING}, \keywordN{GOALS}, \keywordN{STRUCTURE}, \keywordN{SHOWCASE}, and \keywordN{RELEVANCE}. Two researchers conducted open, inductive coding of the responses. Disagreements were resolved through discussion, and codes were refined iteratively. While we did not compute inter-rater reliability, the coding process was guided by grounded theory practices. Table \ref{tab:qualitative-analysis} displays these categories in ascending order by frequency. Except for the last two categories (\keyword{SHOWCASE} and \keywordN{RELEVANCE}), each category appeared in more than one-fifth of the responses as a reason for learning motivation, with the \keyword{VISUAL PROGRESSION} and \keywordN{EFFECTIVE LEARNING} categories peaking at 26\% and 23\%, respectively.
It's important to note that our coding approach allows for the same response to contribute to different categories simultaneously, even within the same GDE. 
Next, we will present the advantages of the various GDEs for the different motivation categories based on the participants' responses.

\textbf{Progress bar} is the most frequently mentioned GDE among those chosen by participants as the most important for increasing their learning motivation. This GDE was primarily associated with the \keyword{VISUAL PROGRESSION} category (n=15). Participants highlighted the progress bar's ability to visually represent their learning progress in an intuitive and immediate manner, which they found highly motivating. There were also mentions related to the \keyword{GOALS} (n=6) and \keyword{OVERVIEW} (n=4) categories. Participants indicated that the progress bar helped them organize their studies by allowing them to estimate the overall content extent, which, in turn, enabled them to set specific learning goals. Here are two responses that illustrate the motivational benefits of progress bars:
\begin{quoteBox}
    ... From a subjective point of view, there is something satisfying about looking at the bar filling up at each step of the way. From a practical point of view, it gives you a visual representation of how much you have done and how much you still need to do. Great to know when you want to organize your study sessions and set goals. [P5]
\end{quoteBox}
\begin{quoteBox}
    For me, progress bars are an essential tool when learning to keep track of how far I have come in a certain topic and overall in a course, which is a motivation in itself ... [P50]
\end{quoteBox}

\textbf{Concept Map} was the second most often mentioned GDE for making learning as engaging as possible, mainly through its contribution to the category of \keyword{STRUCTURE} (n=13). Participants frequently mentioned the benefit of concept maps to provide a comprehensive, clearly structured overview of the learning content. An important motivational potential of concept maps is given by its strength of making connections from the learning content within the big picture visible as for example illustrated by the following responses:
\begin{quoteBox}
    Probably Concept Map, so I can see how everything is connected from what I've learned ... [P30]
\end{quoteBox}
\begin{quoteBox}
    Concept Maps or Storytelling, as they put the topic into perspective (by giving motivation and context) and break it down into digestible pieces [P66]
\end{quoteBox}
Another participant highlighted the motivational potential of concept maps through their use for discovering new topics of interest:
\begin{quoteBox}
    It is especially important for me to see my progress while learning and to discover the different topic areas. That’s why concept maps and progress bars are the most important elements for me. [P35]
\end{quoteBox}

\textbf{Achievements} and \textbf{badges} appear to serve similar motivational purposes for learners. Although badges and achievements were conceptually distinct in our design (e.g., badges as individual task rewards, achievements as milestones), we recognize the overlap may not have been fully clear to participants. To mitigate confusion, each prototype was accompanied by brief contextual examples and presented in randomized order. They are distributed relatively evenly across the \keyword{VISUAL PROGRESSION} and \keywordN{GOALS} categories and are frequently mentioned together (n=6). One prominent effect highlighted by participants is the sense of collecting, which contributes to both making achievements visible and establishing worthwhile goals, ultimately resulting in longer and more engaging learning sessions:
\begin{quoteBox}
    [With badges or achievements] I can easily retrospectively recognize what I've earned, see something I want to work towards ... [P77]
    ... and have the various levels of learning neatly organized in one place. [P77]
\end{quoteBox}
\begin{quoteBox}
    ... During learning I'm glad to have some badges and achievements to collect - they make my learning sessions longer. ... [P62]
\end{quoteBox}
However, participants also emphasized the importance of providing a relevant context for the earned achievements:
\begin{quoteBox}
    Achievements, if related to specific lessons, set out specific goals to work towards and give you bragging rights once you reach them. It also enables a sort of wish to collect them all. [P83]
\end{quoteBox}

\textbf{Feedback} emerged as the most prominent GDE for the EFFECTIVE LEARNING category (n=12). Qualitative analysis highlights the importance of feedback in enhancing learners' motivation by providing effective support throughout the learning process. Participants emphasized that feedback can reduce frustration, maintain learners' focus, and deepen their understanding of the learning context:
\begin{quoteBox}
    The feedback element, since it helps the most in improving, which in turn hopefully lowers the frustration that might stem from making the same mistakes over and over again. [P7]
\end{quoteBox}
\begin{quoteBox}
    Feedback would help me to stay focused and to better understand certain things. [P61]
\end{quoteBox}
Feedback is regarded as essential for learners to acquire the necessary knowledge to contextualize their learning progress:
\begin{quoteBox}
    ... Feedback, so I can test my knowledge and see if I actually learned something ... [P30]
\end{quoteBox}
One participant underscored the distinct significance of feedback compared to other GDEs:
\begin{quoteBox}
    ... However, without any feedback or correction I wouldn't use the app. [P62]
\end{quoteBox}

\textbf{Storytelling} appears to be particularly effective in providing meaningful context for learning content, falling under the \keyword{RELEVANCE} category. This GDE captures learners' attention and keeps them engaged throughout the learning process:
\begin{quoteBox}
    With good storytelling, I can embed everything I've learned into a context. The other elements are important, but they essentially come afterward. With a story, I get engaged initially and the other elements keep me involved then [P34]
\end{quoteBox}
Several participants also highlighted storytelling's advantage in supporting learning strategies that enhance knowledge retention, aligning with the \keyword{EFFECTIVE LEARNING} category:
\begin{quoteBox}
    Storytelling. I believe it will give me a deeper understanding of the subject and it will make it easier for me to remember the knowledge because the story gives me "pictures" of context that are connected and relevant to the knowledge I need to remember. [P13]
\end{quoteBox}

\textbf{Leaderboards} were selected as significant Game Design Elements (GDEs) for enhancing learning motivation by participants who generally consider themselves competitive and enjoy comparing their performance with peers, striving to perform as well as or better than others. One participant emphasized the importance of leaderboards accurately reflecting learning performance:
\begin{quoteBox}
    I like a competitive aspect and therefore Leaderboards based on performance would motivate me most [P48]
\end{quoteBox}
Another participant emphasized the controversial nature of leaderboards, noting that they can be both motivating and demotivating:
\begin{quoteBox}
    Leaderboards can be both motivating and demotivating depending on how competitive one is. It might be cool to have a feature where users can enable or disable this function in the settings. [P47]
\end{quoteBox}

\textbf{Points} are mentioned in a total of ten responses, but rarely the reason for choosing the element was further specified. Most participants simply acknowledged the motivational potential of points without providing specific details:
\begin{quoteBox}
    The points would be a motivational boost [P61]
\end{quoteBox}
Two participants, however, associated points with a sense of progression, emphasizing the motivational aspect of collecting them:
\begin{quoteBox}
    Collecting badges and points - a good feeling of having achieved something. [P76]
\end{quoteBox}
\begin{quoteBox}
    Points. I find it thrilling to try to reach 100 percent [P31]
\end{quoteBox}

In summary, the qualitative analysis revealed six distinct categories of learning motivation, shedding light on how various GDEs serve distinct purposes in enhancing learning motivation (Table \ref{tab:qualitative-analysis}). \textbf{Progress bar} shows to be highly effective in enhancing learning motivation by making learning progress distinctly visible. \textbf{Concept map} excels in providing a structural framework for learning content and revealing connections. Both \textbf{achievements} and \textbf{badges} are instrumental in bolstering motivation by providing clear objectives as well as making success tangible. \textbf{Feedback} contributes to improved learning motivation by enhancing the efficiency of learning. \textbf{Storytelling} offers contextualization to the learning content. \textbf{Leaderboards} stand out for their capacity to furnish a tool for peer comparison and competition.

\section{Discussion and Future Research}
These findings support the need for purpose-driven gamification, that is, gamification strategies that integrate closely with instructional objectives rather than offering extrinsic rewards alone. The motivational themes align well with Self-Determination Theory, particularly the emphasis on competence (via feedback and progress) and autonomy (via personalization).
Our study reveals that among the presented ten GDEs, the progress bar stands out as the \textbf{most effective in enhancing learning motivation}. It is followed by achievements, feedback, concept maps, and points. In contrast, virtual currency and avatars are less favored. 
In a similar study  by Sch{\"o}bel et al. \citep{schoebel2016agony} about user preferences on a different set of ten GDEs in learning management systems the top four preferred GDEs were level, points, goals, and status, which at first may seem different from our results. However, since the meaning of level is closely related to our concept of progress bar and status is closely related to the concept of feedback, the relevance of these two elements seems to be evident, also supported by results from a study by Jia et al. \citep{jia2016personality} about the motivational affordances of GDEs.

The relevance of \textbf{feedback} is further highlighted by the fact, that feedback is considered one of the most powerful tools for influencing learning \citep{hattie2007power}. This is reflected in our findings of the qualitative analysis, where respondents clearly indicated that tools that provide strategies to make learning more effective are essential for their learning motivation. Feedback was mentioned as the most important element to achieve this goal, followed by the GDEs storytelling and concept map.

Our study results indicate \textbf{progress bars} as one of the most central GDEs in increasing learning motivation. Previous research could also demonstrate the motivational potential of progress bars \citep{manzano2021between, ding2017studies, jent2018using}, but still, there is very little research available related to this GDE \citep{mazarakis2023gamification}, and almost no attention is given to it in the context of educational gamification \citep{antonaci2019effects}. This might be due to its nature of appearing very simplistic, but considering the large set of conditional variables such as shape \citep{li2021shape, ohtsubo2014does}, color \citep{hamada2011color}, or general behavior \citep{harrison2007rethinking, conrad2010impact, gronier2019does} that can have an influence on the effectiveness of progress bars, the need for further research about the requirements for progress bars in the design of educational systems is highlighted.

Our findings are consistent with previous research related to \textbf{achievements} and \textbf{badges} and their potential to increase learning motivation \citep{botte2020motivation, hallifax2019factors, reid2015digital}. But it is important to note that the effect strongly depends on how these systems are designed \citep{abramovich2013badges, almeida2023negative, botte2020motivation}. As the results of our qualitative analysis indicate, the main purposes of these elements are on the one hand to contribute to the visualization of progress making learning success tangible, and on the other hand to provide clear goals in the learning process. If achievements and badges are not dosed well enough or become too detached from the contextual purpose (i.e. the learning content), they become more and more extrinsically motivated, which might even have a negative impact on learning motivation \citep{abramovich2013badges, ccubukccu2017gamification}. This demonstrates the necessity of carefully considering the intersection between the design of these elements and the learning content \citep{abramovich2013badges, almeida2023negative, botte2020motivation}.

While \textbf{concept map} was one of the most preferred GDEs by our respondents, studies in this area of gamification research are scarce \citep{klock2020tailored, dehghanzadeh2023gamification}. Our vision of the GDE concept map is any approach to visualizing domain content, its connections, and the learners' progression among different content. Related terms in the literature are for example progression graphs, knowledge maps, semantic maps, domain maps, graph organizers, or advance organizers \citep{ghanizadeh2020graphic, hall2002graphic, klock2020tailored, toda2019analysing}. Our results show that the potential of concept maps to increase learning motivation is especially because of their ability to provide a clear structured overview of the learning topics and their connections. It is also helpful for the learner to orient in the learning context, which has shown to be relevant for enhancing learning attitude and achievement \citep{wahl2011advance, hwang2011interactive, hwang2013concept}. Another benefit of the concept map mentioned by participants was the possibility of exploring new areas where to continue learning. This strategy of activating new content as the learner progresses and showing the next steps proved to be effective in increasing intrinsic motivation \citep{jent2018using, roubi2019towards}. This effect could be even enhanced by utilizing the concept map as a tool to make interdisciplinary connections and everyday life connections visible \citep{tytler2019challenged, ng2020engaging, marquardt2023saving}. 
We suggest that the concept map could be integrated into educational gamification by incorporating existing GDEs like progress bars, feedback, points, or achievements. This integration can take various forms: these elements could be embedded within different sections of the concept map, or progression through the concept map could be rewarded with achievements for example. This approach would ensure a strong connection between the learning content and the gamification system.
In conclusion, we see significant potential in leveraging the benefits of concept maps within the domain of educational gamification. However, it's important to note that further research and evidence are necessary to fully explore these possibilities.
We propose the concept map as a could be implemented in a way that it incorporates existing GDEs such as progress bars, feedback, points, or achievements. On the one hand, those elements could be embedded into different areas of the concept map, and on the other hand, progressing through the concept map could for example be rewarded with achievements. This would ensure the connection between learning content and the gamification system. Overall we believe there are huge possibilities in integrating the advantages of concept maps to the domain of educational gamification, but more research and evidence is essential.

Contrary to our initial expectations, our analysis considering different \textbf{user traits} like gender, video game enjoyment, and age, did not reveal significant differences in preferences for GDEs. These findings align with observations made by Jent and Janneck regarding gender and age preferences in gamified contexts \citep{jent2018using} and also with results of a study by Oliveira et al. that indicated no significant differences between tailored and non-tailored gamification environments \citep{oliveira2020tailoring, oliveira2022effects}.

However, regarding \textbf{age}, our analysis revealed an interesting anomaly among younger participants (age: 15-19). In this age group, the GDE points was significantly more preferred compared to other age groups. Additionally, our results indicated a positive correlation between age and the preference for the GDE concept map, with older participants tending to show a higher preference for this particular GDE compared to younger participants. However, it's important to note that our sample size for the specific age groups, especially the younger ones, was relatively small. Additionally, the study by Jent and Janneck \citep{jent2018using} did not include this age group, limiting direct comparisons. These results underscore the necessity for further research dedicated to learners in primary and secondary education. Such studies should offer more detailed differentiation, considering that expectations, needs, and requirements may vary and change \citep{bruckman2012human}.

Regarding \textbf{gender} differences, our analysis indicates that females exhibit a slightly higher preference for storytelling. This aligns with prior research highlighting gender-related preferences \citep{bentz2022identification}. However, our study did not find significant differences in GDE preferences based on gender overall. It's worth noting that this topic is extensively debated in the literature, with substantial evidence suggesting that gender could indeed influence GDE preferences \citep{manzano2021between, klock2020tailored, pedro2015does}. These findings might imply that the GDEs selected for our study are relatively gender-independent. Alternatively, gender-related differences could manifest during the actual implementation and usage of these elements. In light of these results, the chosen selection of GDEs can serve as a foundational starting point for further investigations. In our future work, we plan to explore additional dimensions of user traits, including age brackets, as well as player types, which show to play a significant role in tailoring gamification effectively \citep{klock2020tailored, santos2021relationship}.

Rather than isolated elements, learners appear to value combinations that reinforce progress and feedback. This has practical implications for educational platforms that allow instructors to bundle or sequence elements.
Most participants chose to combine several GDEs in their learning experiences, with an average of four to five GDEs per bundle.
Our \textbf{combination analysis} revealed that the most preferred GDEs (progress bar, achievements, concept maps, and feedback) are at the same time most often chosen in \textbf{bundles of GDEs}. On average four to five GDEs were chosen for a bundle of GDEs, which is consistent with earlier findings by Sch{\"o}bel et al. \citep{schoebel2016agony}.
Interestingly, while the points-badges-leaderboards (PBL) triad is the most commonly studied combination of GDEs \citep{khaldi2023gamification, dehghanzadeh2023gamification, nadi2022gamification, manzano2021between, subhash2018gamified}, none of the elements from this triad emerged as the top four most preferred GDEs in our study. The PBL combination was chosen a total of 13 times, with the first occurrence observed in bundles comprising seven or more elements. In contrast, the combination of progress bar, concept map, and feedback was selected 23 times, indicating a greater demand for such combinations.

The preceding observations collectively suggest a correlation between learning satisfaction and learning motivation. This connection becomes evident as the primary rationale of participants for favoring particular GDEs with the aim of enhancing learning motivation lies in their capacity to deliver visual progression, employ effective learning strategies, establish clear objectives, and provide structural support. 
This underscores the importance of expanding research beyond the PBL triad and exploring \textquote[0]{deeper gamification integrating different dynamics, mechanics, and aesthetics that can act as reinforcers of intrinsic motivation of the students} \citep[p. 10]{manzano2021between}. Based on our findings, we recommend initiating an educational gamification design under the consideration of the GDEs progress bar, concept map, achievements, and feedback. Subsequently, other elements like badges, storytelling, points, or leaderboards can be incorporated. In the design and implementation process, careful consideration should be given to the purpose of each element in fulfilling specific learning motivation requirements, as indicated in our qualitative analysis (Table \ref{tab:qualitative-analysis}).
We propose a novel approach to educational gamification that strongly emphasizes a purpose-driven methodology, where GDEs are selected based on their relevance to the learning content and their potential to enhance learning motivation. By establishing this strong connection between design and purpose, it becomes possible to mitigate potential negative effects of gamification in educational contexts such as extrinsic motivation overshadowing learners' intrinsic drive \citep{almeida2023negative}.

Our study highlights the significance of including progress bar, concept map, achievements, feedback, badges, and storytelling in educational gamification. We strongly recommend promoting the interoperability of these elements within educational systems, creating opportunities for innovative gamification approaches that simultaneously engage and effectively support learning processes.

\section{Conclusion}
This study examined learners’ preferences for ten game design elements (GDEs) in educational contexts through a best-worst scaling (BWS) survey and qualitative analysis. Progress bars, achievements, immediate feedback, and concept maps emerged as the most motivating elements—both individually and when combined. These elements were consistently favored for their ability to visualize progress, structure learning, and provide meaningful feedback, demonstrating a clear preference for learning-aligned over reward-centric gamification.

Interestingly, while overall preferences were consistent across demographic groups, younger learners (ages 15–19) showed a stronger preference for points, while affinity for concept maps increased with age. These differences suggest opportunities for age-sensitive or personalized gamification strategies.

Our qualitative analysis identified six motivational drivers that underpin learners’ choices, including visible progress, clarity of goals, and relevance to the learning content. These findings align with Self-Determination Theory and reinforce the importance of supporting intrinsic motivation through the careful integration of GDEs into learning design. It became evident that the most motivating GDEs were those that seamlessly extended the learning experience, making learning progress visible, providing clear goals, facilitating connections, and offering guidance for knowledge expansion. 


Rather than isolated elements, learners appear to value combinations that reinforce progress and feedback. This has practical implications for educational platforms and learning management systems (LMSs) that allow instructors to bundle or sequence GDEs. Purpose-aligned gamification strategies should emphasize elements that directly support learning trajectories, rather than relying solely on extrinsic incentives.
This underscores the importance of future research in exploring novel GDE compositions, including innovative GDEs such as concept maps, which can effectively establish intrinsic connections with the learning content rather than solely relying on extrinsically motivated arbitrary goals.

We consider this work a step toward a paradigm shift in educational gamification, one that is purpose-driven, learner-centered, and grounded in motivational theory. Future research should investigate how these preferences affect actual learning behaviors and outcomes in real-world classroom and digital environments.

\backmatter








\section*{Declarations}


\subsection*{Funding}
This research did not receive any specific grant from funding agencies in the public, commercial, or not-for-profit sectors.

\subsection*{Conflict of Interest}
The authors declare that they have no conflicts of interest or competing interests.

\subsection*{Ethics Approval and Consent to Participate}
Participation in the study was voluntary, anonymous, and complied with applicable institutional and national research ethics standards. As the study involved no personal or sensitive data and presented no foreseeable risks, formal ethics approval was not required under the guidelines of the corresponding institution.

\subsection*{Consent for Publication}
Not applicable. No identifiable data from participants are published.

\subsection*{Data Availability}
An anonymized version of the dataset supporting the findings of this study is available from the corresponding author upon reasonable request.

\subsection*{Materials Availability}
The visual prototypes and survey materials used in the study are available upon request from the corresponding author.

\subsection*{Code Availability}
No custom software code was developed for this study.

\subsection*{Author Contributions}


\bigskip
\begin{flushleft}%
Editorial Policies for:

\bigskip\noindent
Springer journals and proceedings: \url{https://www.springer.com/gp/editorial-policies}

\bigskip\noindent
Nature Portfolio journals: \url{https://www.nature.com/nature-research/editorial-policies}

\bigskip\noindent
\textit{Scientific Reports}: \url{https://www.nature.com/srep/journal-policies/editorial-policies}

\bigskip\noindent
BMC journals: \url{https://www.biomedcentral.com/getpublished/editorial-policies}
\end{flushleft}


\bibliography{references}
\clearpage

\clearpage
\begin{appendices}
\section{Appendices}
\subsection{Prototypes of Game Design Elements}
\label{appendix:prototypes}

\begin{table}[h]
\centering
\resizebox{\linewidth}{!}{%
\begin{tabular}{|l|l|l|}
\toprule
\makecell[l]{\textbf{Avatar} \\
An avatar represents you and is \\
displayed to other app users. You \\
can customize it e.g. appearance, \\
clothes and accessories.} &
\makecell[l]{\textbf{Achievements} \\
In the app you can see an overview \\
with the achievements you can \\
reach. Every achievement can be \\
rewarded by e.g. a badge.} &
\makecell[l]{\textbf{Leaderboard}\\
The leaderboard allows you to \\
compare your learning progress \\
with your friends or other app \\
users.} \\

\includegraphics[width=3cm]{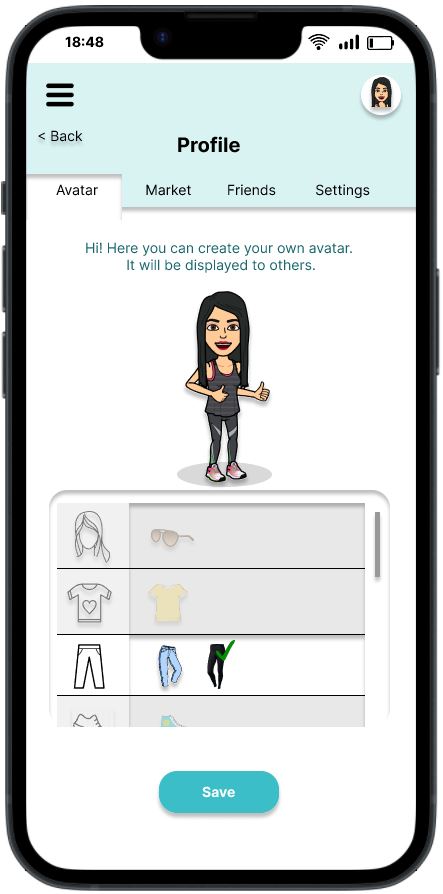} & \includegraphics[width=3cm]{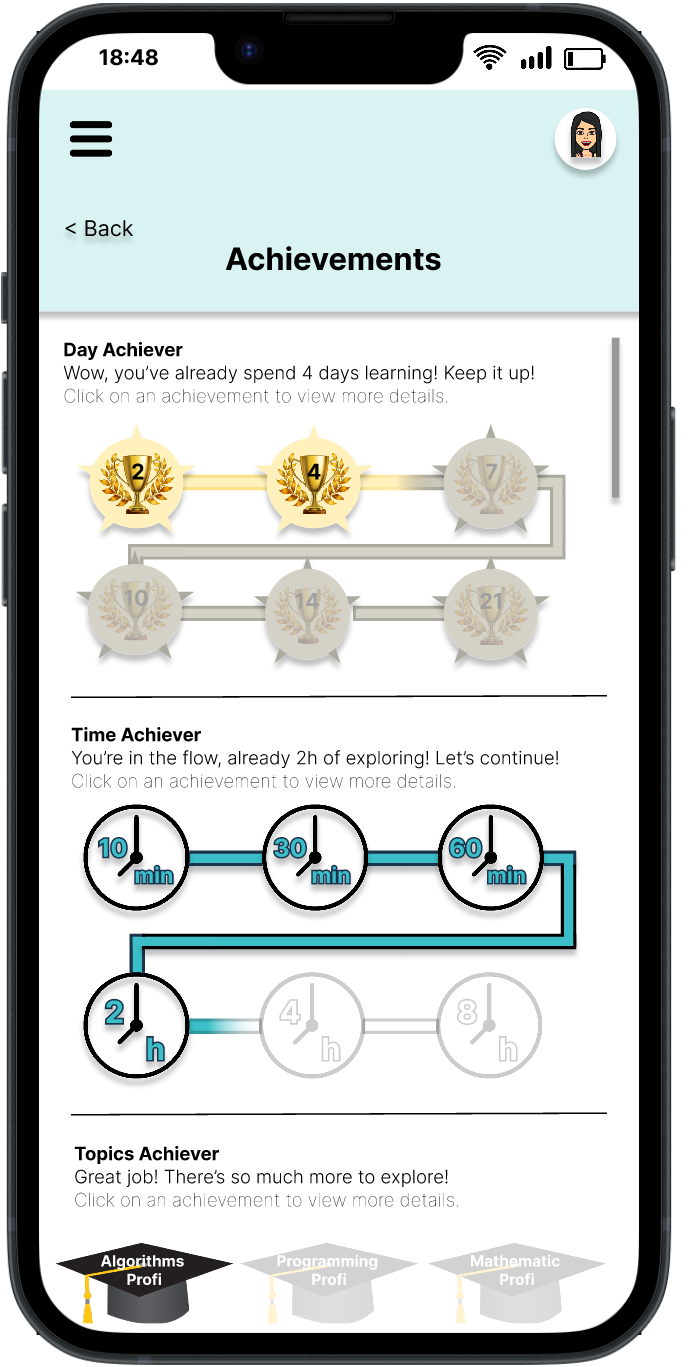} & \includegraphics[width=3cm]{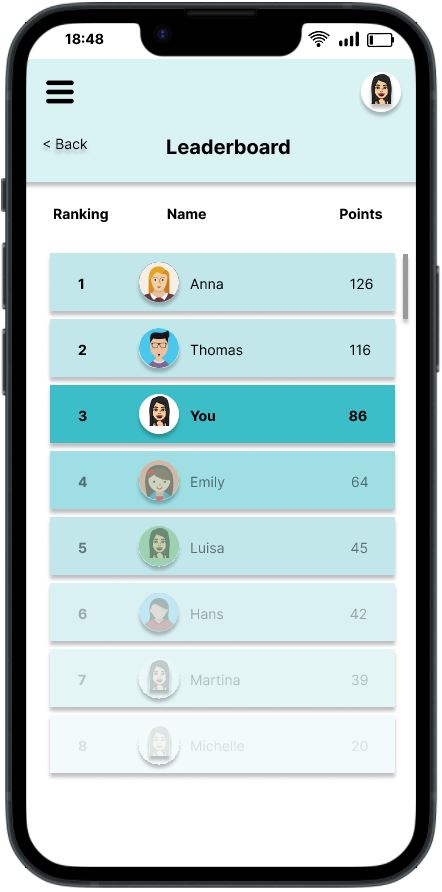}\\

\midrule

\makecell[l]{\textbf{Badges} \\
After finishing a few tasks or \\
challenges like learning for 7 \\
days straight, you can gain badges. \\
You can see all badges in the \\
achievement overview.} &
\makecell[l]{\textbf{Storytelling} \\
In the app you are guided through \\
a story, a kind of mission, in \\
which you can do tasks to solve a \\
certain problem and thus deepen \\
learning content in a playful way.} &
\makecell[l]{\textbf{Feedback} \\
After each completed task, you can \\
toggle feedback by clicking on the \\
work assistant button. The feedback \\
can help you to understand the \\
solutions and give hints.} \\

\includegraphics[width=3cm]{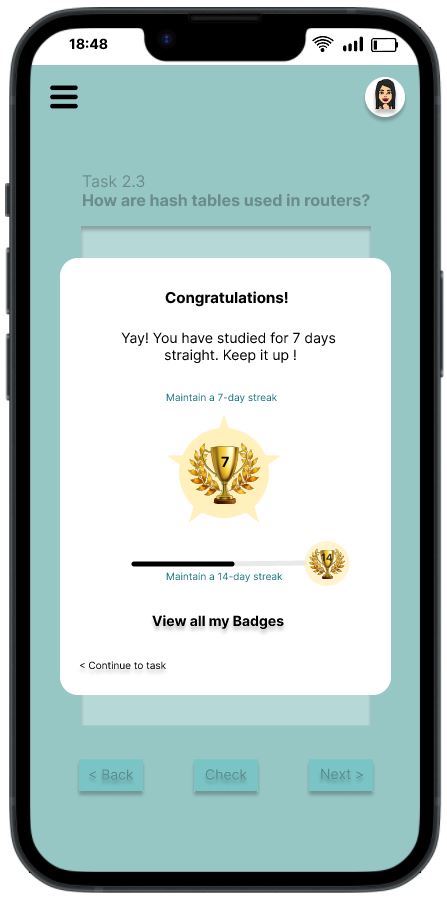} & \includegraphics[width=3cm]{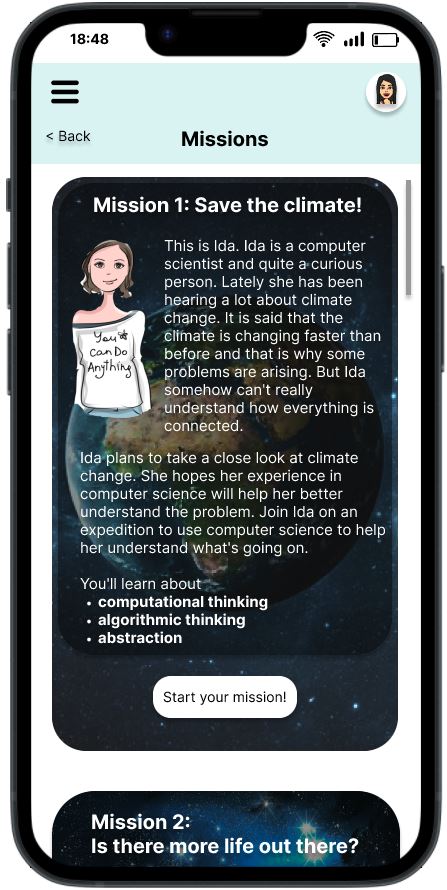} & \includegraphics[width=3cm]{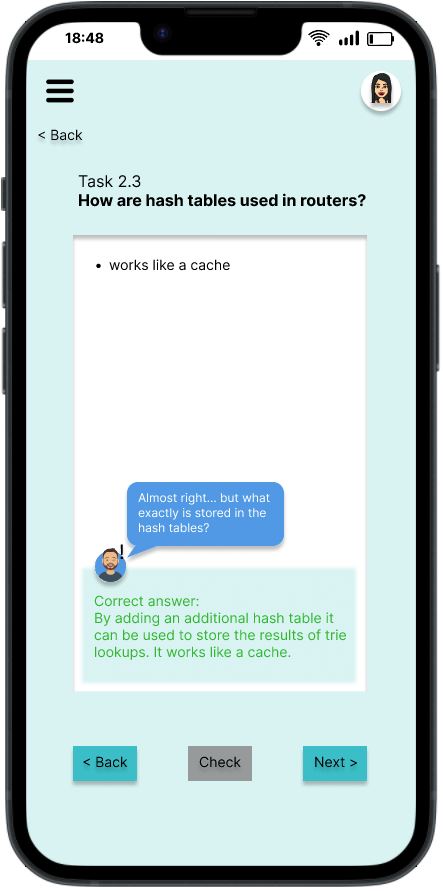}\\

\bottomrule

\end{tabular}
}
\label{t_combination}
\end{table}

\begin{table}[]
\centering
\resizebox{\linewidth}{!}{%
\begin{tabular}{|l|l|l|}
\toprule
\makecell[l]{\textbf{Virtual Currency} \\
Through various completed tasks, you \\
can get virtual currency, which will \\
be added to your wallet. With this \\
currency you can buy different things, \\
for example clothes and accessoires \\
for your avatar for different layouts \\
for the app.} &
\makecell[l]{\textbf{Concept Map} \\
Each subject area is displayed in their \\
respective sub-areas in a concept map. \\
Here you can see all the topics \\
included in the app and their \\
connections to each other. Furthermore, \\
the progress you have made in \\
each area is shown there.} &
\makecell[l]{\textbf{Progress bar} \\
Progress bars show the relative \\
progression of your own learning \\
process. You can see an overall \\
progress while you are working \\
on a task and if you click on it, you \\
can see detailed information broken \\
down by different topics.} \\

\includegraphics[width=3cm]{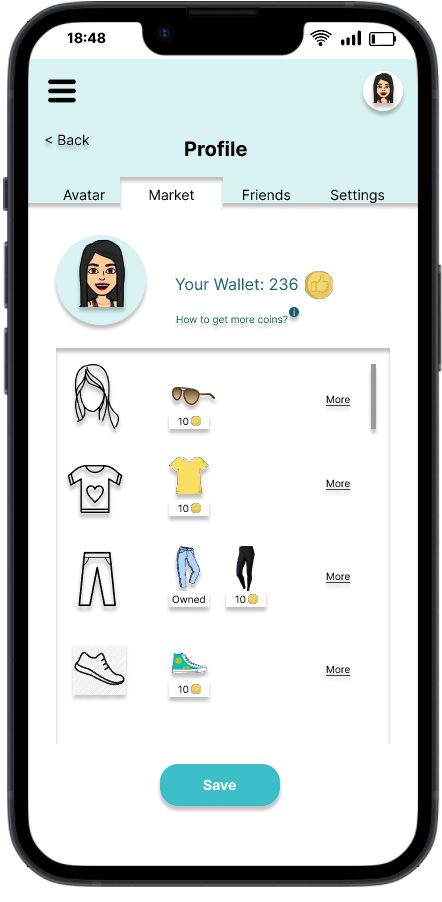} & \includegraphics[width=3cm]{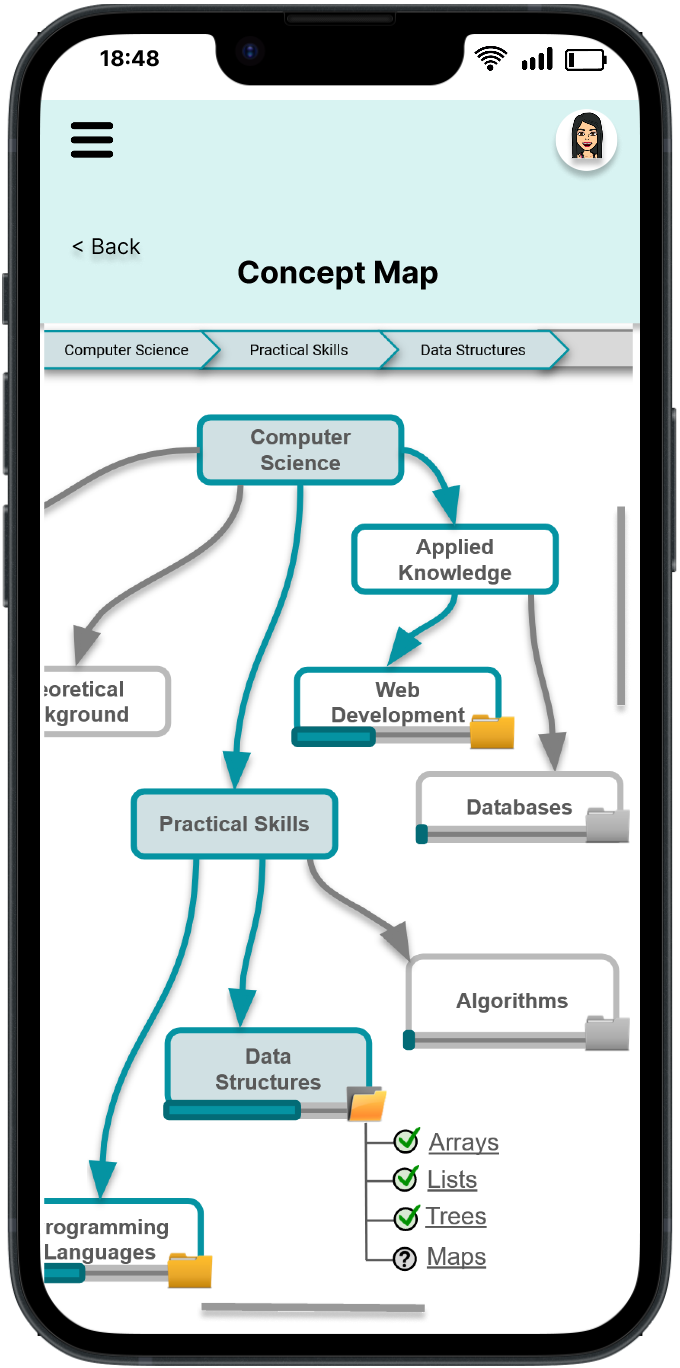} & \includegraphics[width=3cm]{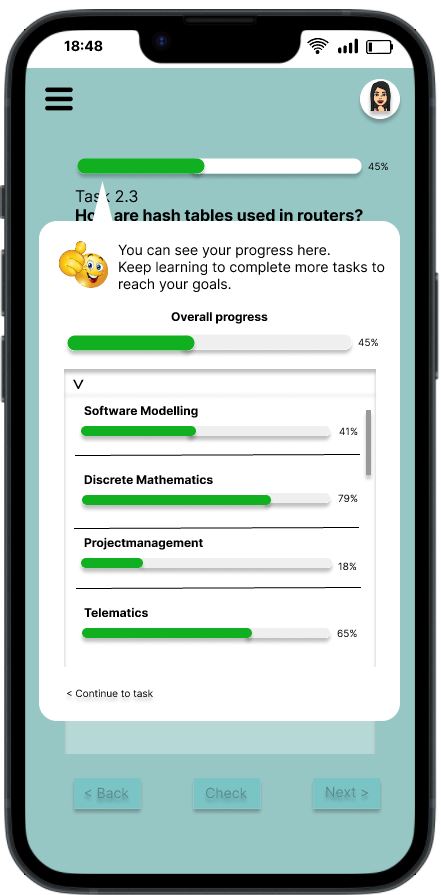}\\

\midrule

\makecell[l]{\textbf{Points} \\
You can earn knowledge-points by \\
activities in the app. You have a \\
total points system that expands after \\
each point you gain. Additionally, by \\
clicking on the information button next \\
to the point notification, you can view \\
detailed information how the total points \\
are put together.} & & \\

\includegraphics[width=3cm]{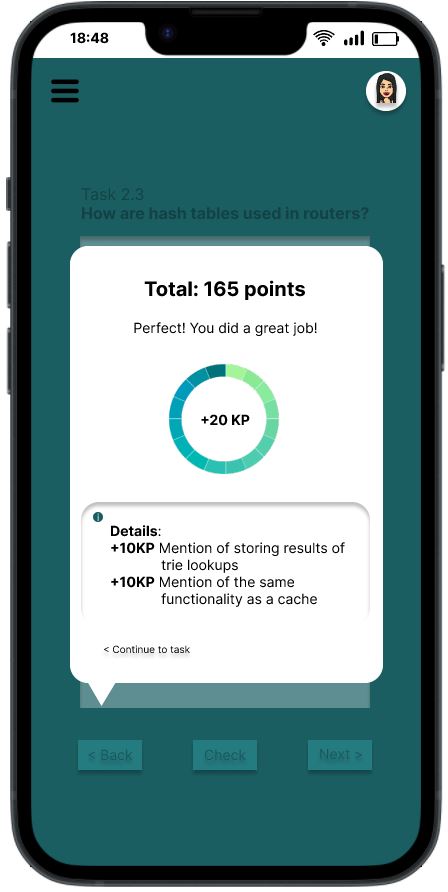} & & \\

\bottomrule

\end{tabular}
}
\label{t_combination}
\caption{Table with description and visual prototypes for each of the ten selected GDEs. Each prototype is presented on a smartphone layout.}
\end{table}

\clearpage
\subsection{Choice Sets}
\label{appendix:choicesets}

\begin{table}[h]
\centering
\resizebox{\linewidth}{!}{%
\begin{tabular}{lcccccccccccccccc}
\toprule
    \multirow{2}{*}{Game Design Element} & \multicolumn{15}{c}{Choice Set ID} & \multirow{2}{*}{Total Appearances} \\
    & 1 & 2 & 3 & 4 & 5 & 6 & 7 & 8 & 9 & 10 & 11 & 12 & 13 & 14 & 15 \\
\midrule
    Progress Bar & & &                       \checkmark & &            \checkmark & & &                       \checkmark & \checkmark & & &                       \checkmark & \checkmark & & & 6 \\
    Feedback & & &                       \checkmark & \checkmark & & & & & &                                                        \checkmark & \checkmark & &            \checkmark & &             \checkmark & 6 \\
    Achievements & &            \checkmark & & & & &                                             \checkmark & \checkmark & \checkmark & \checkmark & & & & &                                              \checkmark & 6 \\
    Virtual Currency & &            \checkmark & &            \checkmark & \checkmark & &            \checkmark & & & &                                  \checkmark & \checkmark & & & & 6 \\
    Leaderboard & & & & &                                             \checkmark & \checkmark & &            \checkmark & &            \checkmark & \checkmark & & &                       \checkmark & & 6 \\
    Avatar & \checkmark & &            \checkmark & &            \checkmark & &            \checkmark & & & & & & &                                                                    \checkmark &  \checkmark & 6  \\
    Points & \checkmark & \checkmark & & & & & & & &                                                                              \checkmark & &            \checkmark & \checkmark & \checkmark & & 6 \\
    Storytelling & &            \checkmark & \checkmark & \checkmark & &            \checkmark & & &                       \checkmark & & & & &                                             \checkmark & & 6 \\
    Conceot Map & \checkmark & & &                       \checkmark & &            \checkmark & \checkmark & \checkmark & & & & &                                             \checkmark & & & 6 \\
    Badges & \checkmark & & & & &                                             \checkmark & & &                       \checkmark & &             \checkmark & \checkmark & & &                       \checkmark & 6 \\
\bottomrule
\end{tabular}
}
\label{t_choicesets}
\end{table}

\end{appendices}

\end{document}